\documentstyle[12pt,aasms4,epsf]{article}
\def\tempest%
{\begin{array}{ccc}
1 & 1 & 1 \\
1 & 1 & 1 \\
4 & 3 & 8
\end{array}}

\begin{document}
\title{The Applicability of the Astrometric Method in Determining\\
       the Physical Parameters of Gravitational Microlenses}
\bigskip
\bigskip

\author{Cheongho Han}
\smallskip
\affil{Department of Astronomy \& Space Science, \\
       Chungbuk National University, Chongju, Korea 361-763 \\
       cheongho@astro-3.chungbuk.ac.kr}
\bigskip
\bigskip
\author{Kyongae Chang}
\smallskip
\affil{Department of Physics, \\
       Chongju University, Chongju, Korea 360-764 \\
       kchang@alpha94.chongju.ac.kr}

\authoremail{cheongho@astro-3.chungbuk.ac.kr}
\authoremail{kchang@alpha94.chongju.ac.kr}
\bigskip

\begin{abstract}
   In this paper, we investigate the applicability of the astrometric method 
to the determination of the lens parameters for gravitational microlensing 
events toward both the LMC and the Galactic bulge.  For this analysis, we 
investigate the dependency of the astrometrically determined angular 
Einstein ring radius, $\Delta(\theta_{\rm E}/\theta_{\rm E,0})$, 
on the lens parameters by testing various types of events.
In addition, by computing $\Delta(\theta_{\rm E}/\theta_{\rm E,0})$
for events with lensing parameters which are the most probable for a given
lens mass under the standard models of Galactic matter density and velocity
distributions, we determine the expected distribution of the
uncertainties as a function of lens mass.
From this study, we find that the values of the angular Einstein ring
radius are expected to be measured with uncertainties
$\Delta(\theta_{\rm E}/\theta_{\rm E,0}) \lesssim 10\%$
up to a lens mass of $M\sim 0.1\ M_\odot$ for both Galactic
disk-bulge and halo-LMC events with a moderate observational strategy.
The uncertainties are relatively large for Galactic bulge-bulge
self-lensing events, $\Delta(\theta_{\rm E}/\theta_{\rm E,0}) \sim 25\%$
for $M\sim 0.1\ M_\odot$, but they can be substantially reduced by
adopting more aggressive observational strategies.  We also find that
although astrometric observations can be performed for most 
photometrically detected Galactic bulge events, a significant fraction 
($\sim 45\%$) of LMC events cannot be astrometrically observed due to 
the faintness of their source stars.

\end{abstract}

\vskip25mm

\centerline{submitted to {\it Monthly Notices of the Royal 
Astronomy Society}: Sep 26, 1998}
\centerline{Preprint: CNU-A\&SS-07/98}
\clearpage

\section{Introduction}

Current microlensing experiments (Alcock et al.\ 1997a, 1997b; Ansari et al.\ 
1996; Udalski et al.\ 1997; Alard \& Guibert 1997)\markcite{alcock1997a,
alcock1997b, ansari1996, udalski1997, alard1997} are detecting massive 
astrophysical compact objects (MACHOs) by monitoring the 
light variations of stars undergoing gravitational microlensing.
The light curve of a microlensing event is related to the lensing 
parameters by
$$
A = { u^2 + 2\over u(u^2+4)^{1/2}};\qquad 
u = \left[\beta^2+\left( {t-t_0\over t_{\rm E}}\right)^2\right]^{1/2},
\eqno(1.1)
$$
where A is the magnification, $u$ is the lens-source separation, 
$\beta$ is the impact parameter,
$t_0$ is the time of maximum amplification, and $t_{\rm E}$ is the 
Einstein ring radius crossing time (Einstein timescale).
Of these lensing parameters, only the timescale provides information 
about the lens because it is the only one directly related to the 
physical parameters of the lens (lens parameters) by 
$$
t_{\rm E} = {r_{\rm E}\over v};\qquad
r_{\rm E} = \left( {4GM\over c^2} {D_{ol}D_{ls}\over D_{os}}\right)^{1/2},
\eqno(1.2)
$$
where $v$ is the lens-source transverse speed, $r_{\rm E}$ is the 
physical size of the Einstein ring radius, $M$ is the mass of the lens, 
and $D_{ol}$, $D_{ls}$, and $D_{os}$ are the separations between the 
observer, lens, and source star.
However, due to the degeneracy in the lens parameters, 
the nature of the lens population is still very uncertain.

   The uncertainties in the lens parameters can be significantly reduced
if one can measure the lens proper motion, $\mu=\theta_{\rm E}/t_{\rm E}$,
where $\theta_{\rm E}=r_{\rm E}/D_{ol}$ is the angular Einstein ring radius.
The lens proper motions can be measured both photometrically 
(Gould 1994; Nemiroff \& Wickramasinghe 1994; Witt \& Mao 1994)\markcite{
gould1994, nemiroff1994, witt1994} and spectroscopically (Maoz \& Gould 1994; 
Loeb \& Sasselov 1995).\markcite{maoz1994, loeb1994}
However, measuring the proper motion with either of these techniques 
is only possible for several special classes of lensing events.

   The lens proper motions can be measured in a general way if the 
separation between the two source star images can be measured.
As this image separation is generally on order $1\ \mu{\rm as}$, 
it can not be measured with current instrumentation. 
However, the proposed {\it Space Interferometry Mission} 
(Allen, Shao, \& Peterson 1997, hereafter SIM)\markcite{allen1997}  
will have high enough positional accuracy to measure
the astrometric displacement in the light centroid
caused by gravitational microlensing (Walker 1995; H\o\hskip-1pt g, 
Novikov \& Polnarev 1995; Paczy\'nski 1998; Boden, Shao, \& 
Van Buren 1998).\markcite{walker1995, hog1995 paczynski1998, boden1998}
The astrometric shift of the source star image centroid is related to the 
lensing parameters by
$$
\vec{\delta\theta} = {\theta_{\rm E}\over u^2+2} 
\left( {\cal T}\hat{\bf x} + \beta\hat{\bf y} \right),
\eqno(1.3)
$$
where ${\cal T}=(t-t_0)/t_{\rm E}$ and the $x$ and $y$ axes represent the 
directions which are parallel and normal to the lens-source transverse motion, 
respectively.
The trajectory of the apparent source star image traces out an ellipse 
during the event (astrometric ellipse, Jeong, Han, \& 
Park 1998).\markcite{jeong1998}  With the measured astrometric ellipse, 
one can determine the impact parameter of the event because the shape (i.e. 
the axis ratio) of the ellipse is related to the impact parameter $\beta$.
The importance of measuring the astrometric centroid shifts is that one can 
then determine the angular Einstein ring radius as the 
size of the astrometric ellipse, i.e., the semimajor axis, is directly 
proportional to $\theta_{\rm E}$.  Once the angular Einstein ring radius 
is determined from the astrometric shifts, one can determine the lens 
proper motion from $\theta_{\rm E}$   combined with the event timescale, 
which is determined from photometric monitoring of the event.

   Since astrometric centroid shifts of microlensing events exhibit 
various shapes 
depending on the combination of the lensing parameters, i.e., 
$\beta$, $t_{\rm E}$, and $\theta_{\rm E}$, the uncertainties in the
astrometrically determined values of the angular Einstein ring radius,
$\Delta(\theta_{\rm E}/\theta_{\rm E,0})$, will depend on the 
lensing parameters.  The values of the lensing parameters $t_{\rm E}$ and 
$\theta_{\rm E}$ are related to the physical parameters of lenses, 
$D_{ol}$, $v$, and $M$, 
and thus the uncertainties will also depend on the lens parameters.
In addition, due to the differences in the lens parameters for different 
types of lensing events, the expected values of 
$\Delta(\theta_{\rm E}/\theta_{\rm E,0})$ for LMC events will be 
different from those of Galactic bulge events.  As a result, the 
expected uncertainties of $\theta_{\rm E}$ will have a wide range. 

   Boden, Shao, \& Van Buren (1998)\markcite{boden1998} performed a 
detailed study of the performance of the astrometric observations of 
gravitational microlensing events in determining various observables, 
including $\theta_{\rm E}$.
From this analysis, they demonstrated the usefulness of astrometric 
observations of lensing events in determining the lens parameters.   
In their analysis, however, Boden et al.\ (1998)\markcite{boden1998}
considered only LMC events, while the majority of events are detected toward 
the Galactic bulge.  In addition, their estimate of the uncertainties is 
based on the very limited number of test cases they considered.
For example, all of the events they considered had a fixed lens 
location of $D_{ol}=8\ {\rm kpc}$,  and their only values for the Einstein 
timescale and impact parameter were $t_{\rm E}=0.1$ and 0.2 yr and 
$\beta=0.4$ and 0.8, respectively.  They also did not take into account 
the restriction the source star brightness can impose on astrometric 
observations.  Therefore, one cannot make general conclusions about the 
usefulness of astrometrically monitoring lensing events based 
on their analysis alone.

   In this paper, we investigate the applicability of the astrometric 
method to the determination of the lens parameters for events toward both 
the LMC and the Galactic bulge.  For this analysis, we investigate the 
dependency of $\Delta(\theta_{\rm E}/\theta_{\rm E,0})$ on the lens parameters 
by testing various types of events.
In addition, by computing $\Delta(\theta_{\rm E}/\theta_{\rm E,0})$
for events with the most probable lensing parameters for a given 
lens mass under the standard models of the Galactic matter density and velocity 
distributions, we determine the expected distribution of the 
uncertainties in $\theta_{\rm E}$ as a function of a lens mass. 
From this study, we find that the values of the angular Einstein ring 
radius are expected to be measured with uncertainties 
$\Delta(\theta_{\rm E}/\theta_{\rm E,0}) \lesssim 10\%$ 
up to the lens mass of $M\sim 0.1\ M_\odot$ for both Galactic 
disk-bulge and halo-LMC events with a moderate observational strategy. 
The uncertainties are relatively large for Galactic bulge-bulge 
self-lensing events, $\Delta(\theta_{\rm E}/\theta_{\rm E,0}) \sim 25\%$ 
for $M\sim 0.1\ M_\odot$, but they can be substantially reduced by 
adopting more aggressive observational strategies.  We also find that 
although astrometric observations can be performed for most photometrically 
detected Galactic bulge events, a significant fraction ($\sim 45\%$) of 
LMC events cannot be astrometrically observed due to the faintness of 
their source stars.

\section{Dependency of Lens Parameters}
   
   To show the applicability of the astrometric method to the determination
of $\theta_{\rm E}$, we begin our analysis by investigating how the 
uncertainties in an astrometrically determined $\theta_{\rm E}$ changes 
with respect to varying lens parameters. 
   We determine the uncertainties in $\theta_{\rm E}$  
by conducting model fits to the astrometric shifts of simulated 
events with various lensing parameters.
The result of the fit is obtained by computing $\chi^2$, i.e., 
$$
\chi^2 = {1\over \sigma_{\delta\theta_c}^2} 
\sum_{i=1}^{N_{\rm obs}} \left\{
\left[\delta\theta_{c,x}(t_i)-\delta\theta_{c0,x}(t_i)\right]^2 +
\left[\delta\theta_{c,y}(t_i)-\delta\theta_{c0,y}(t_i)\right]^2 \right\},
\eqno(2.1)
$$
where $N_{\rm obs}$ is the number of the astrometric observations 
and $(\delta\theta_{c0,x},\delta\theta_{c0,y})$ and 
$(\delta\theta_{c,x},\delta\theta_{c,y})$ are the centroid 
shifts of the simulated and model events, respectively.
Following the mission specifications of SIM, we adopt the positional 
accuracy of the astrometric measurements to be 
$\sigma_{\delta\theta_c}=0.01\ {\rm mas}$ (http://huey.jpl.nasa.gov/sim).
Since our goal is to see the variation of 
$\Delta(\theta_{\rm E}/\theta_{\rm E,0})$ with respect to individual lensing 
parameters, the uncertainties are determined under fixed but 
realistic observational conditions.  We will discuss the expected uncertainties 
under other observational conditions in \S\ 4.
To determine the uncertainties, we assume that the astrometric 
centroid shifts of events are measured with a frequency $f=1\ {\rm day}^{-1}$ 
during $t_{\rm min}\leq t_{\rm obs}\leq t_{\rm max}$, where $t_{\rm min}$ 
and $t_{\rm max}$ are the times of the first and last astrometric 
measurements.  Since the astrometric observations will be performed for 
events that are photometrically detected, we set $t_{\rm min}=-0.4t_{\rm E}$
while $t_{\rm max}=3.0t_{\rm E}$.
The uncertainties are determined by the $3\sigma$ level, which is
equivalent to $\Delta\chi^2=\chi^2-\chi_{\rm min}^2=9$, where 
$\chi_{\rm min}^2$ is the best-fitting $\chi^2$ value.
To illustrate the sensitivity of this uncertainty on the lens parameters, we
have tested a sample event with lensing parameters  
$(\beta, t_{\rm E}, \theta_{\rm E}) = 
(0.5, 11.3\ {\rm days}, 0.22\ {\rm mas})$.  Then the dependency of the 
uncertainty on each lensing parameter is obtained by varying the parameter 
of interest while holding the other lensing parameters constant.

   In Figure 1, we present the uncertainties of the astrometrically determined 
angular Einstein ring radius with respect to various lensing parameters.
From the figure, one finds that the uncertainties increase significantly 
with decreasing angular Einstein ring radius and decreasing Einstein timescale.
On the other hand, the dependency of 
$\Delta(\theta_{\rm E}/\theta_{{\rm E},0})$ on the impact parameter is 
negligible.

\section{Dependency on Lens Masses}

   In the previous section we showed that the uncertainty 
$\Delta(\theta_{\rm E}/\theta_{\rm E,0})$ depends strongly on the size of 
the angular Einstein ring radius and the duration of the Einstein timescale.
Since both $\theta_{\rm E}$ and $t_{\rm E}$ are related to the mass of the 
lens, the strong dependency of $\Delta(\theta_{\rm E}/\theta_{{\rm E},0})$ 
on $\theta_{\rm E}$ and $t_{\rm E}$ implies that the uncertainties 
for events caused by different populations of lenses will be greatly different.
Then the naturally arising questions are: ``Is the astrometric observation
of lensing events a guaranteed method of determining $\theta_{\rm E}$?" and 
if not, ``To what extent can one determine $\theta_{\rm E}$ with this method?''
To answer these questions, one must determine the expected value of 
$\Delta(\theta_{\rm E}/\theta_{{\rm E},0})$ as a function of lens mass.

   We determine the relation between $\Delta(\theta_{\rm E}/\theta_{\rm E,0})$
and the lens mass by computing the uncertainties for events with lensing 
parameters which are the most probable for a given lens mass.
Because the values of $\theta_{\rm E}$ and $t_{\rm E}$ also depend on the  
lens parameters $v$ and $D_{ol}$, there is no one-to-one correspondence 
between the lens mass and the lensing parameters.
However, with models of the Galactic matter density and velocity distributions,
one can statistically determine the most probable values of the lensing 
parameters corresponding to individual lens masses. 

   With models for the Galactic matter density and velocity distributions, 
the distributions of $t_{\rm E}$ and $\theta_{\rm E}$ for a given
lens mass $M$ are obtained by
$$
f(t_{\rm E}) = 
\int_{0}^\infty dD_{os}\rho(D_{os})
\int_0^{D_{os}} dD_{ol}\rho(D_{ol}) \pi r_{\rm E}^2
\int_0^\infty \int_0^\infty dv_y dv_z
v f(v_y,v_z)
$$
$$
\delta\left[ t_{\rm E}- 
\left( {4GM\over c^2v^2}{D_{ol}D_{ls}\over D_{os}}\right)^{1/2}
\right],
\eqno(3.1)
$$
for the Einstein timescale and
$$
f(\theta_{\rm E}) =
\int_{0}^\infty dD_{os}\rho(D_{os})
\int_0^{D_{os}} dD_{ol}\rho(D_{ol}) \pi r_{\rm E}^2
\int_0^\infty \int_0^\infty dv_y dv_z
v f(v_y,v_z)
$$
$$
\delta\left[ \theta_{\rm E}- 
\left( {4GM\over c^2}{D_{ls}\over D_{ol}D_{os}}\right)^{1/2}
\right],
\eqno(3.2)
$$
for the angular Einstein ring radius.
Here $\rho(D_{ol})$ and $\rho(D_{os})$ are the density distributions
of lens and source stars, $\delta$ is the delta function, 
$v_y$ and $v_z$ are the components of the transverse velocity ${\bf v}$, 
and $f(v_y,v_z)$ is their distribution.
In the above equations, the factors $\pi r_{\rm E}^2$ and $v$ are included 
because events with larger cross-sections and higher transverse speeds
are more likely to occur.
For Galactic bulge events (disk-bulge plus bulge-bulge self-lensing
events), $v_y$ and $v_z$ represent the velocity components which are normal 
and azimuthal to the Galactic plane, respectively.
On the other hand, since we adopt a non-rotating isotropic velocity model
for LMC events (halo-LMC events, see Table 1), these variables represent 
two velocity components of arbitrary direction in the plane normal 
to the line of sight toward the LMC.
   We assume that the individual components of the transverse velocities 
have Gaussian distributions of the form
$$
f(v_i)\propto \exp 
\left[ -{ (v_i-\bar{v}_i)^2\over 2\sigma_{v_{i}}^2}\right],\qquad
i\in \{ y, z\}.
\eqno(3.3)
$$
The values of the mean, $\bar{v}_i$, and the dispersion, $\sigma^2_{v_i}$,
of the velocity distributions, which are listed in Table 1, are adopted from 
the models of Han \& Gould (1995)\markcite{han1995} for the Galactic bulge 
events and from Han \& Gould (1996)\markcite{han1996} for the LMC events.  
For the matter density distributions of the individual Galactic components, 
we adopt an axisymmetric model (Kent 1992)\markcite{kent1992} for the Galactic 
bulge, a double-exponential disk model (Bahcall 1986)\markcite{bahcall1986} 
for the Galactic disk, and an isothermal sphere model with a core radius 
(Bahcall, Schmidt, \& Soneira 1983)\markcite{bahcall1983} for the Galactic 
halo.  The adopted models of the matter density distributions are listed in 
Table 2.  The adopted distances to the Galactic center and to the LMC are
$8.0\ {\rm kpc}$ and $55\ {\rm kpc}$, respectively.

   In Figure 2, we present the expected distribution of $f(\theta_{\rm E})$ 
and $f(t_{\rm E})$ for various types of events and different values of lens 
mass.  
From the figure, one can see that the expected distribution for a given 
type of event covers a wide range, considering the variety of lens 
populations and corresponding lens masses.
In addition, even with a fixed lens mass, the distributions for
different types of events are substantially different from each other.
Therefore, the analysis of $\Delta(\theta_{\rm E}/\theta_{{\rm E},0})$
based on limited sets of lensing parameters for 
only a single type of lensing events will lead to erroneous 
conclusions about the general applicability of the astrometric method
to the determination of $\theta_{\rm E}$.

   Once the most probable values of $\theta_{\rm E}$ and $t_{\rm E}$ for 
a given lens mass are obtained from their distributions, the expected
values of $\Delta(\theta_{\rm E}/\theta_{\rm E})$ are determined as 
before, by carrying out a $\chi^2$ fit to the astrometric centroid shifts 
of the event with the most probable lensing parameters.
Since the dependency of the uncertainties on the impact parameter
is not important, we assume $\beta=0.5$ for all simulated events.
The astrometric centroid shifts of the events are assumed to be 
measured under moderate observational conditions 
with a frequency $f=0.5\ {\rm day}^{-1}$ during 
$-0.4t_{\rm E}\leq t_{\rm obs}\leq 3.0t_{\rm E}$ and with a positional 
accuracy of $\sigma_{\delta\theta_{c}}=0.01\ {\rm mas}$.
Then, the relation between $\Delta(\theta_{\rm E}/\theta_{\rm E})$ 
and lens mass is obtained by repeating the uncertainty determinations
for different lens masses.

   In Figure 3, we present the resulting relation between the lens mass 
and the expected uncertainty in the astrometrically determined values of 
$\theta_{\rm E}$.  We find that the angular Einstein ring radii are
expected to be measured with uncertainties
$\Delta(\theta_{\rm E}/\theta_{\rm E,0}) \lesssim 10\%$ up to the lens 
mass of $M\sim 0.1\ M_\odot$ for both Galactic disk-bulge and halo-LMC events.
For Galactic bulge-bulge self-lensing events, the uncertainty is relatively
large: $\Delta(\theta_{\rm E}/\theta_{\rm E,0}) \sim 25\%$ for
$M\sim 0.1\ M_\odot$.

\section{Improved Astrometric Observational Strategies}

   Up to now, we have investigated the uncertainties 
$\Delta(\theta_{\rm E}/\theta_{{\rm E},0})$ for a fixed observational 
strategy.
In this section, we investigate the dependency of the uncertainties on 
observational strategies to find the optimal astrometric observational 
strategies for better determinations of $\theta_{\rm E}$. 
The accuracy of the astrometric determination of $\theta_{\rm E}$ 
can be broadly improved in three ways.
First of all, the uncertainties will be decreased if the astrometric positional 
accuracy can be improved.  However, since the positional accuracy is restricted 
by the instrumentation and not by the observational strategy, we do 
not consider this scenario further. Secondly, the accuracy can be improved 
by increasing the duration of the astrometric observations.
Finally, increasing the frequency of astrometric observation will also help 
to improve the accuracy in determining $\theta_{\rm E}$.

   In Figure 4, we present the uncertainties as functions of the frequency
(upper panel) and the duration (lower panel) of the astrometric measurements 
of $\delta\theta_c$.  The uncertainties are obtained for an example event 
with a set of lensing parameters $(\beta, t_{\rm E}, \theta_{\rm E})=
(0.5, 11.3\ {\rm days}, 0.22\ {\rm mas})$ under various observational 
conditions.  From the figure, one finds that the accuracy improves 
significantly with increasing observational frequencies.
Longer observations of the event also contribute to decrease the uncertainty 
of $\theta_{\rm E}$ determination.  However, observations longer than 
$t_{\rm max}\sim 3t_{\rm E}$ do not help to a significant reduction of
the uncertainty.

   As we know that measuring $\delta\theta_c$ with increased frequency
will help to improve the accuracy in determining $\theta_{\rm E}$, we 
re-determined the relation between $\Delta(\theta_{\rm E}/\theta_{{\rm E},0})$
and the lens mass for a new observational strategy with 
$f=3\ {\rm day}^{-1}$.
Since we already tested the uncertainties for a long enough observational 
duration of $t_{\rm max}=3t_{\rm E}$, we only increased the observational 
frequency for this test.
In Figure 2, we present the newly determined 
$\Delta(\theta_{\rm E}/\theta_{{\rm E},0})$-$M$ relation (represented by 
a thick solid line) for the Galactic bulge-bulge self-lensing events, 
which have the largest uncertainties among the various types of events.
One finds that the accuracy of the $\theta_{\rm E}$ determination significantly 
improves -- the uncertainty decreases by approximately one-half over
the entire range of lens masses.

\section{Restriction by Source Star Brightness}

   The applicability of the astrometric method in determining lens parameters
is additionally restricted by source star brightness because the astrometric 
observations will only be possible for events with bright source stars.
For both the Galactic bulge and LMC fields, the photometric monitoring 
of source stars is restricted by crowding.  Current microlensing experiments 
toward the Galactic bulge reach the photometric detection limit when the 
stellar density of the fields arrives at $\sim 10^4\ {\rm stars}\ 
{\rm deg}^{-2}$ (C. Alcock 1997, private communication).
Based on the model luminosity function of Han (1997)\markcite{han1997}, 
this corresponds to $V\sim 20.5$, which approximately coincides with the 
astrometric detection limit of SIM ($V\sim 20$, http://huey.jpl.nasa.gov/sim).
Therefore, the astrometric observations of microlensing events can be 
achieved for most of the photometrically detected Galactic bulge events.
By using the $R$-band luminosity function provided by Gould (1998)\markcite{
gould1998}, we also determine the fraction of LMC events with source stars 
bright enough for astrometric observation.
With the photometric determination limit of $V\sim 21$ (Alcock et al.\ 
1997b)\markcite{alcock1997b}, which corresponds to $R\sim 21.1$, we find 
that nearly half ($\sim 45\%$) of LMC source stars are fainter than the 
astrometric detection limit ($R\sim 20.25$).
This implies that for LMC events the major restriction in the general
applicability of the astrometric $\theta_{\rm E}$ determination comes 
from the faintness of the source stars and not from the large uncertainties in 
the determined values of $\theta_{\rm E}$.

\section{Conclusion}

   In this paper, we have considered the effectiveness of astrometric
monitoring of gravitational microlensing events in determining the lens 
parameters. The main results from this analysis can be summarized as follows: 
 
\begin{enumerate}
\item
The uncertainties of the astrometrically determined angular Einstein ring 
radii are strongly dependent on the values of $\theta_{\rm E}$ and $t_{\rm E}$ 
for the events.  However, the dependency of 
$\Delta(\theta_{\rm E}/\theta_{{\rm E},0})$ on the impact parameter 
is not important.
\item
The distributions of $\theta_{\rm E}$ and $t_{\rm E}$ for events caused 
by various populations of lenses cover wide ranges.
Even with a fixed lens mass, the distributions for different
types of lensing events are substantially different from each other.
Therefore, the analysis of $\Delta(\theta_{\rm E}/\theta_{{\rm E},0})$ 
based on limited sets of lensing parameters for only a single type of 
lensing events will lead to erroneous conclusions about the general
applicability of the astrometric method in determining lens parameters.
\item
Measurements of $\delta\theta_c$ with increased frequencies during longer
times of observations help to improve the accuracy of $\theta_{\rm E}$
determination.  However, measurements longer than $t_{\rm max}\sim 3t_{\rm E}$ 
do not contribute to a significant reduction in 
$\Delta(\theta_{\rm E}/\theta_{{\rm E},0})$.
\item
With a moderate observational strategy, the value of $\theta_{\rm E}$ 
can be determined with an uncertainty 
$\Delta(\theta_{\rm E}/\theta_{{\rm E},0})\lesssim 10\%$ up to a lens mass 
of $M\sim 0.1\ M_\odot$ for the Galactic disk-bulge and halo-LMC events.
The uncertainties for the Galactic bulge-bulge self-lensing events is 
relatively large with $\Delta(\theta_{\rm E}/\theta_{{\rm E},0})\sim
25\%$ for $M\sim 0.1\ M_\odot$.
However, the uncertainty can be reduced substantially by adopting a more 
aggressive observational strategy.
\item
Astrometric observations will be possible for most photometrically detected
Galactic bulge events.  However, nearly half of LMC events will not be 
astrometrically observable due to the faintness of their source stars.
Therefore, for LMC events the major restriction in the general applicability
of the astrometric determination of $\theta_{\rm E}$ comes from the source 
star brightness and not from the large uncertainties in the determined 
value of $\theta_{\rm E}$.
\end{enumerate}
    
\acknowledgements
We would like to thank P.\ Martini for careful reading the manuscript.
We also would like to acknowledge the support of the Korea Research 
Foundation for the 1999 program year.

\clearpage

\begin{center}
\bigskip
\bigskip
\centerline{\small {TABLE 1}}
\smallskip
\centerline{\small {\sc The Transverse Velocity Distribution Models}}
\smallskip
\begin{tabular}{ccccc}
\hline
\hline
\multicolumn{1}{c}{event type} &
\multicolumn{1}{c}{$\bar{v}_y$} &
\multicolumn{1}{c}{$\sigma_{v_y}$} &
\multicolumn{1}{c}{$\bar{v}_z$} &
\multicolumn{1}{c}{$\sigma_{v_z}$} \\
\multicolumn{1}{c}{} &
\multicolumn{1}{c}{$({\rm km\ s}^{-1})$} &
\multicolumn{1}{c}{$({\rm km\ s}^{-1})$} &
\multicolumn{1}{c}{$({\rm km\ s}^{-1})$} &
\multicolumn{1}{c}{$({\rm km\ s}^{-1})$} \\
\hline
bulge-bulge  & $-220(1-D)$ & $[100^2(1+D)^2]^{1/2}$  & 0 & $[100^2(1+D)^2]^{1/2}$ \\
disk-bulge   & $220D$      & $[30^2+100^2D^2]^{1/2}$ & 0 & $[20^2+100^2D^2]^{1/2}$ \\
halo-LMC     & 0           & $250/\sqrt{2}$          & 0 & $250/\sqrt{2}$ \\
\hline
\end{tabular}
\end{center}
\smallskip
\noindent
{\footnotesize \qquad NOTE.---
The transverse velocity distribution models for different types of microlensing 
events.  The distributions are assumed to be Gaussian in form, 
$f(v_i)=\exp\left[ -(v_i-\bar{v}_i)^2/2\sigma^2_{v_i}\right],\ i\in \{ y,z\}$,
and the values of the mean and the standard deviation of the 
distributions are listed.  Here $D=D_{ol}/D_{os}$.
}

\vskip2cm
\begin{center}
\bigskip
\bigskip
\centerline{\small {TABLE 2}}
\smallskip
\centerline{\small {\sc The Model Matter Density Distributions}}
\smallskip
\begin{tabular}{cl}
\hline
\hline
\multicolumn{1}{c}{Galactic } &
\multicolumn{1}{c}{distribution} \\
\multicolumn{1}{c}{components} &
\multicolumn{1}{c}{($M_\odot\ {\rm pc}^{-3}$)} \\
\hline
disk  & $\rho(R,z)=0.06\exp\left\{ -\left[ (R-R_0)/3500+z/325\right]\right\}$ \\
bulge & $\rho(s) = 1.04\times 10^6(s/0.482)^{-1.85}$ (for $s< 938\ {\rm pc}$) \\ 
      & $\rho(s) = 3.53K_0 (s/667)$                   (for $s\geq 938\ {\rm pc}$) \\
halo  & $\rho(r) = 7.9\times 10^{-3} (r_c^2 + R_0^2)/(r_c^2+r^2)$ \\
\hline
\end{tabular}
\end{center}
\smallskip
\noindent
{\footnotesize \qquad NOTE.---
In the Galactic bulge model, $s^4 = R^4+(z/0.61)^4$, $R=(x^2+y^2)^{1/2}$ 
where $x$ and $z$ represent the axes directed along the line of sight and
toward the Galactic pole.
The notation $K_0$ represents a modified Bessel function.
The adopted values of the solar Galactocentric distance and 
the core radius of the halo is $R_0=8\ {\rm kpc}$ and
$r_c=2\ {\rm kpc}$, respectively.
}

\begin{figure}[t]
\centerline{\hbox{\epsfysize=16truecm \epsfbox{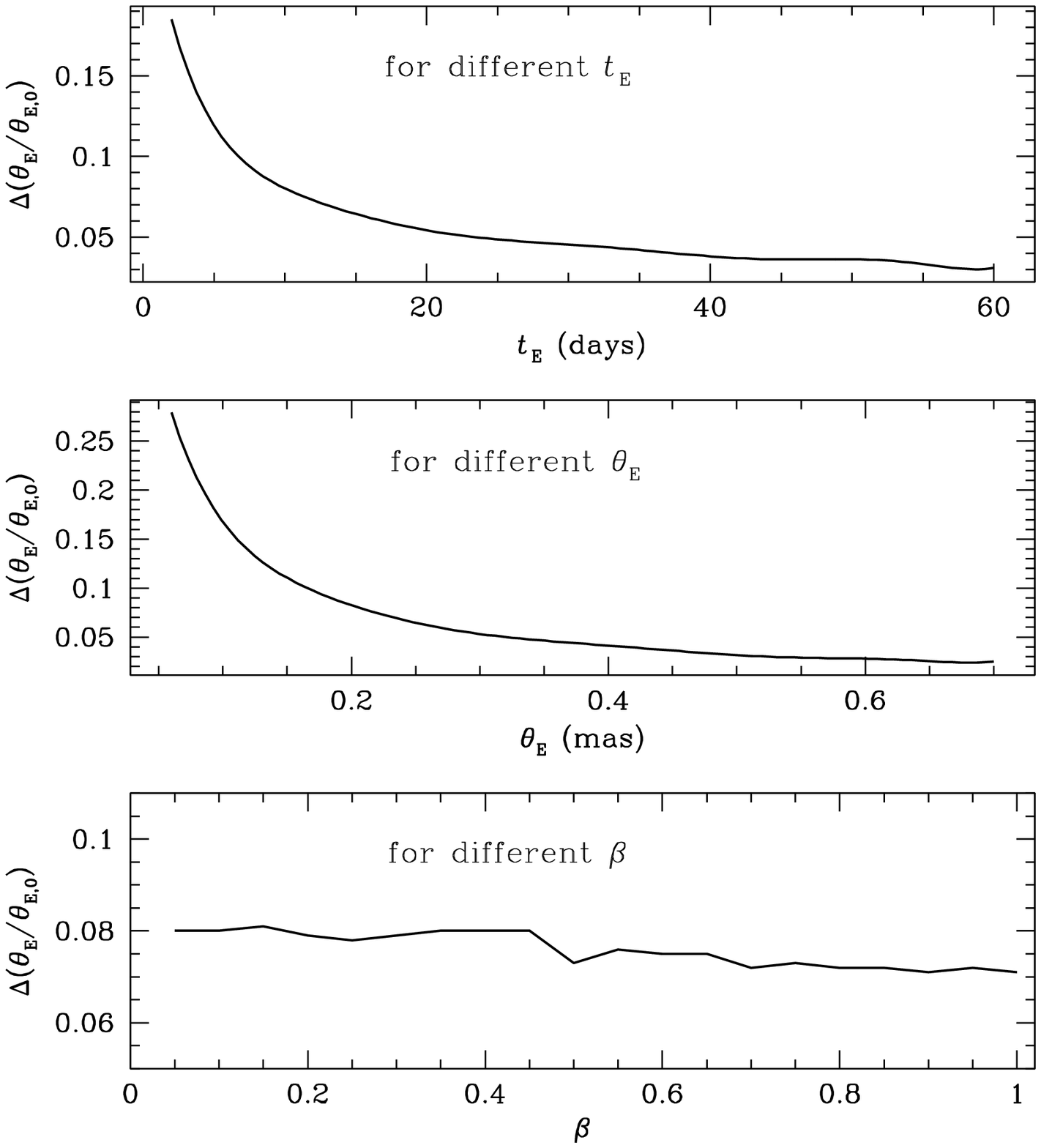}}}
\noindent
{\small {\bf Figure 1:}\
The dependency of the uncertainties of the astrometrically determined 
$\theta_{\rm E}$ on various lensing parameters.
The astrometric centroid shifts of the event are assumed 
to be measured with a frequency $f=1\ {\rm day}^{-1}$ during
$-0.4t_{\rm E}\leq t_{\rm obs}\leq 3.0t_{\rm E}$ and with a positional 
accuracy $\sigma_{\delta\theta_c}=0.01\ {\rm mas}$.
To investigate the dependencies, we test an example event with a set
of lensing parameters of $(\beta,t_{\rm E},\theta_{\rm E})
=(0.5, 11.3\ {\rm days}, 0.22\ {\rm mas})$.  The dependency of the 
uncertainties on each lensing parameter is obtained by varying the 
parameter of interest while holding the other lens parameters constant.
One finds that the uncertainties $\Delta(\theta_{\rm E}/\theta_{E,0})$ 
increase significantly with decreasing angular Einstein ring radius and 
with decreasing Einstein timescale.  On the other hand, the dependency of 
$\Delta(\theta_{\rm E}/\theta_{E,0})$ on the impact parameter is negligible.
}
\end{figure}

\begin{figure}[t]
\centerline{\hbox{\epsfysize=16truecm \epsfbox{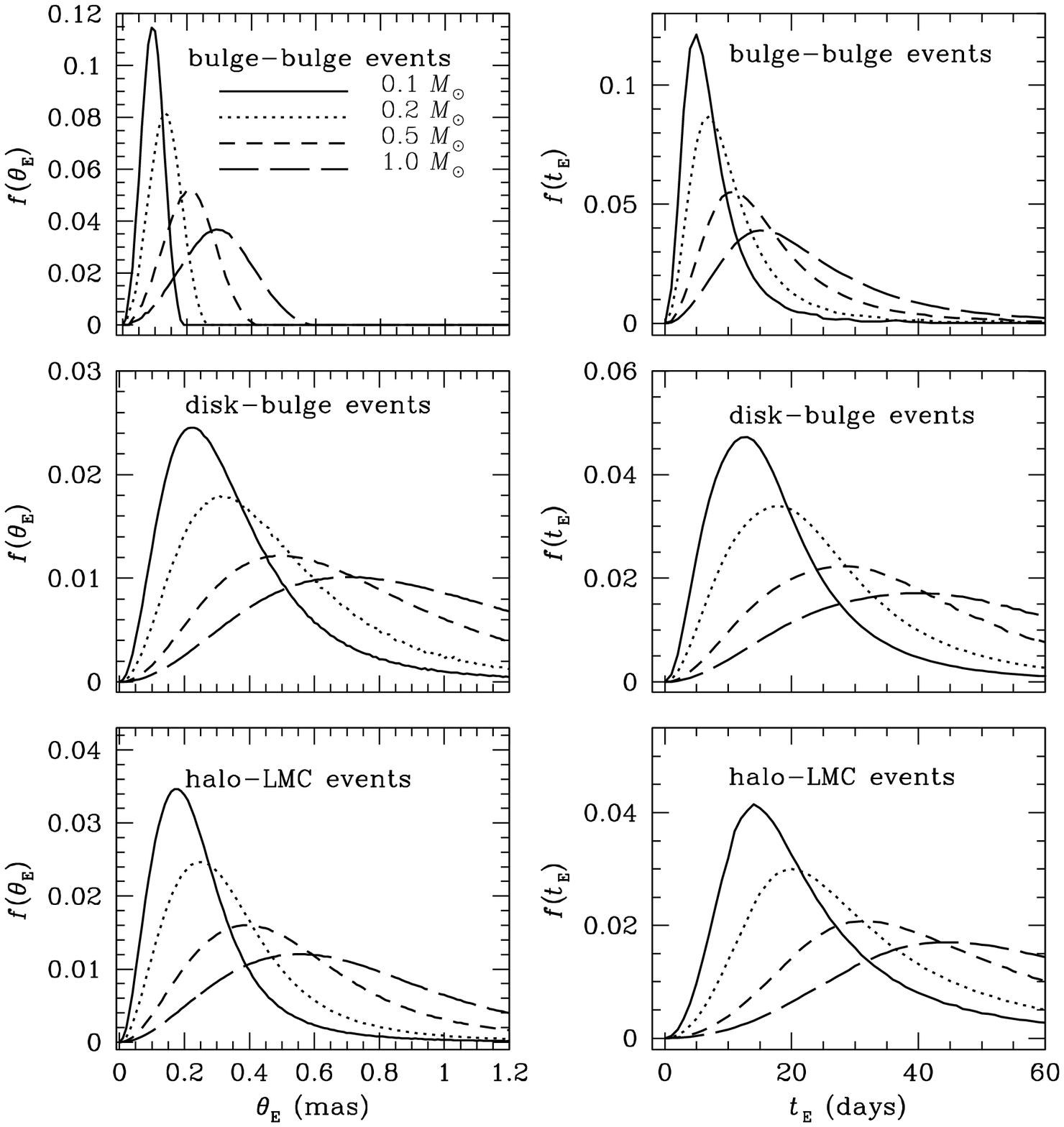}}}
\noindent
{\small {\bf Figure 2:}\
The distributions of the Einstein ring radii and the Einstein timescales
for various types of events and different values of lens mass.  
From the figure, one finds that the expected distribution for a given 
type of event covers a wide range, considering the variety of lens populations
and corresponding lens masses.  In addition, even with a fixed lens mass, 
the distributions for different types of events are substantially different 
from each other.  Therefore, the analysis of 
$\Delta(\theta_{\rm R}/\theta_{\rm E,p})$ based on limited sets of lensing 
parameters for only a single population of lensing events will lead to 
erroneous conclusions about the general applicability of the astrometric 
method to the determination of $\theta_{\rm E}$.
}
\end{figure}

\begin{figure}[t]
\centerline{\hbox{\epsfysize=16truecm \epsfbox{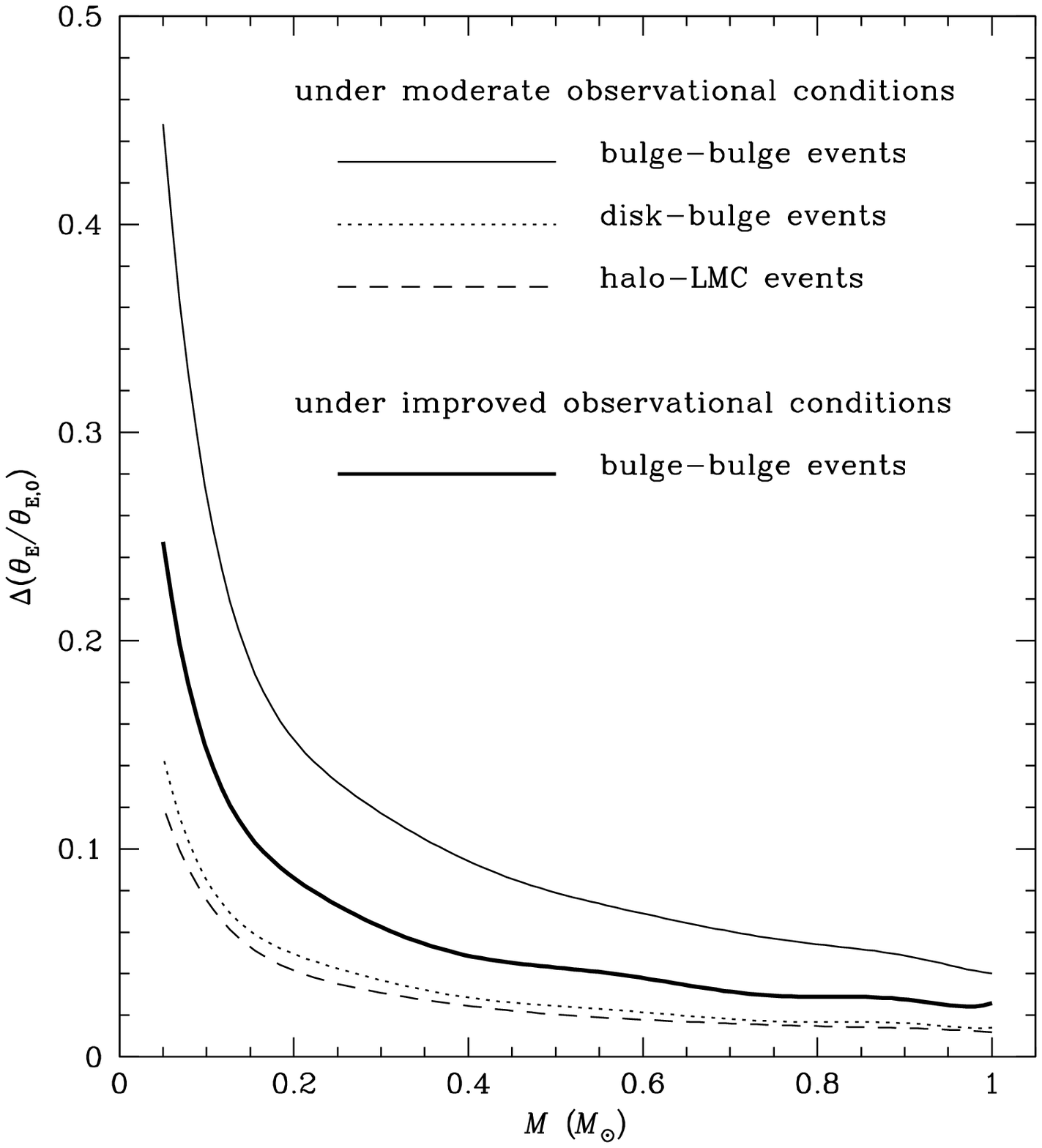}}}
\noindent
{\small {\bf Figure 3:}\
The relation between the lens mass and the expected uncertainties in 
the astrometrically determined values of angular Einstein ring radii
for various types of events.  The uncertainties are determined by
carrying out $\chi^2$ fits to the astrometric centroid shifts of the 
events with lensing parameters $\theta_{\rm E}$ and $t_{\rm E}$, 
which are the most probable for a given lens mass under the models of 
Galactic matter density and velocity distributions.
Since the dependency of the uncertainty on the impact parameters
is not important, we assume $\beta=0.5$ for all events.
For the moderate observational conditions, the astrometric centroid 
shifts of the events are assumed to be measured with $f=0.5\ {\rm day}^{-1}$ 
during $-0.4t_{\rm E}\leq t_{\rm obs}\leq 3.0t_{\rm E}$ and a positional 
accuracy of $\sigma_{\delta\theta_c}= 0.01\ {\rm mas}$.
The $\Delta(\theta_{\rm R}/\theta_{\rm E,p})$-$M$ relation for the 
Galactic bulge events is re-determined under a new observational strategy 
with $f=3\ {\rm day}^{-1}$ and it is represented by a thick solid line.
}
\end{figure}

\begin{figure}[t]
\centerline{\hbox{\epsfysize=16truecm \epsfbox{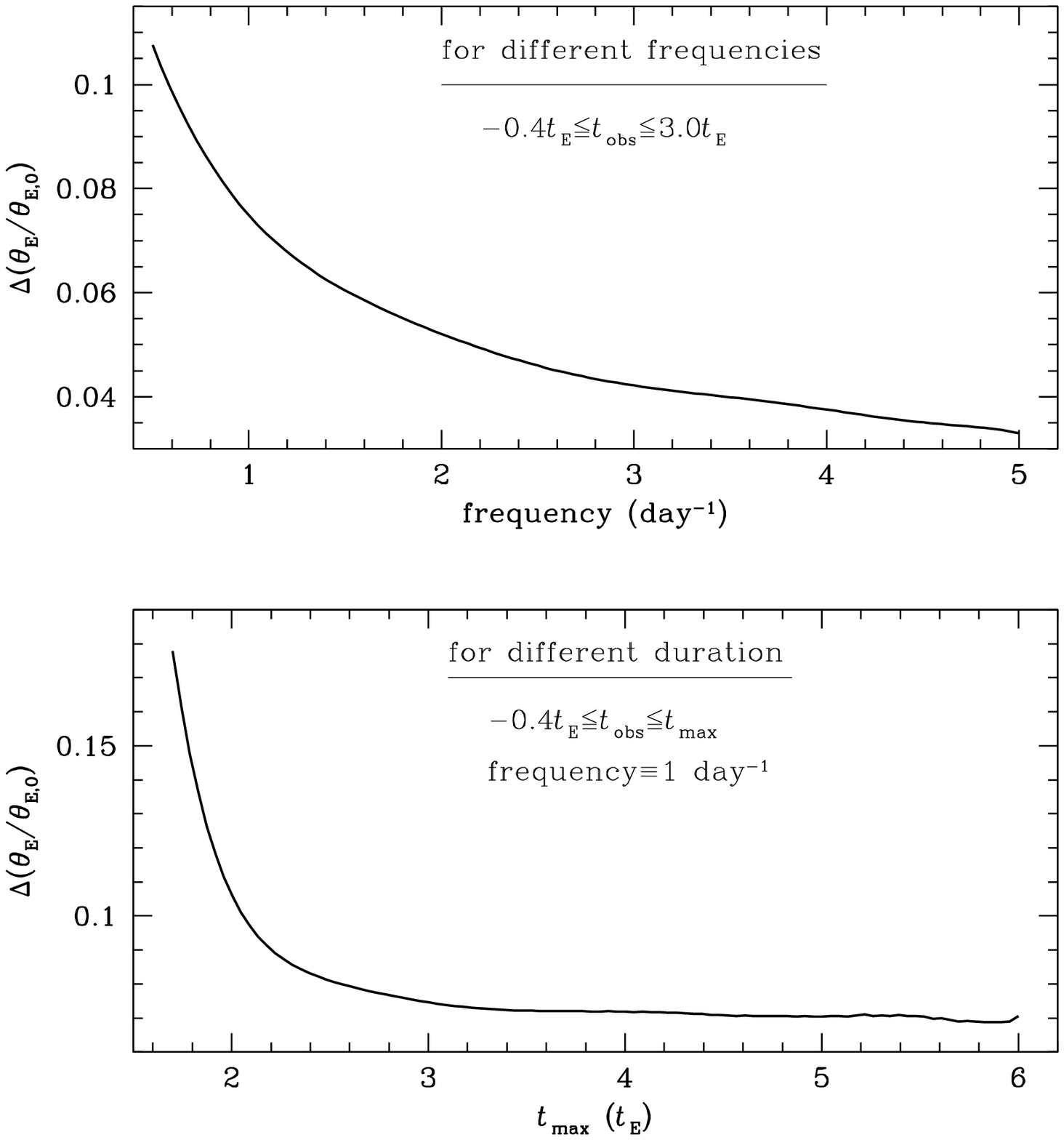}}}
\noindent
{\small {\bf Figure 4:}\
The dependency of the astrometrically determined $\theta_{\rm E}$ 
on the observational frequencies (upper panel) and the
duration of measurements (lower panel).  The uncertainties are determined 
for an example event with a set of lensing parameter of
$(\beta,t_{\rm E},\theta_{\rm E})=(0.5, 11.3\ {\rm days}, 0.22\ {\rm mas})$
under different frequencies and durations of astrometric measurements of 
the centroid shifts.  From the figure, one finds that the accuracy of 
$\theta_{\rm E}$ determination improves significantly with increasing 
observational frequencies.  Longer observations of the event also contribute
to decrease the uncertainty of $\theta_{\rm E}$ determination.
However, measurements of $\delta\theta_{c}$ longer than 
$t_{\rm max}\sim 3t_{\rm E}$ do not help to a significant reduction of the 
uncertainty.
}
\end{figure}

\end{document}